# Nanoscale transient magnetization gratings excited and probed by femtosecond extreme ultraviolet pulses

D. Ksenzov[1*], A. A. Maznev[2], V. Unni[3], F. Bencivenga[4], F. Capotondi[4], A. Caretta[4], L. Foglia[4], M. Malvestuto[4], C. Masciovecchio[4], R. Mincigrucci[4], K. A. Nelson[2], M. Pancaldi[3], E. Pedersoli[4], L. Randolph[1], H. Rahmann[1], S. Urazhdin[5], S. Bonetti[3,6], C. Gutt[1]

[1] *Department Physik, Universität Siegen, Walter-Flex-Str. 3, 57072, Siegen, Germany*

[2] *Department of Chemistry, Massachusetts Institute of Technology, 77 Massachusetts Avenue, Cambridge, MA 02139, USA.*

[3] *Department of Physics, Stockholm University, 106 91 Stockholm, Sweden*

[4] *Elettra Sincrotrone Trieste S.C.p.A., Strada Statale 14, km 163.5, 34149 Basovizza (TS), Italy.*

[5] *Department of Physics, Emory University, Atlanta, Georgia 30322, USA*

[6] *Department of Molecular Sciences and Nanosystems, Ca' Foscari University of Venice, 30172 Venice, Italy*

*Corresponding author: *dmitriy.ksenzov@uni-siegen.de*

**Abstract**

We utilize coherent femtosecond extreme ultraviolet (EUV) pulses derived from a free electron laser (FEL) to generate transient periodic magnetization patterns with periods as short as 44 nm. Combining spatially periodic excitation with resonant probing at the dichroic M-edge of cobalt allows us to create and probe transient gratings of electronic and magnetic excitations in a CoGd alloy. In a demagnetized sample, we observe an electronic excitation with 50 fs rise time close to the FEL pulse duration and ~0.5 ps decay time within the range for the electron-phonon relaxation in metals. When the experiment is performed on a sample magnetized to saturation in an external field, we observe a magnetization grating, which appears on a sub-picosecond time scale as the sample is demagnetized at the maxima of the EUV intensity and then decays on the time scale of tens of picoseconds via thermal diffusion. The described approach opens prospects for studying dynamics of ultrafast magnetic phenomena on nanometer length scales.

**Introduction**

Since the first observation of laser-induced ultrafast demagnetization by Beaurepaire et al. (*1*), interest in optical control of magnetization on ultrafast time scales has been continuously increasing, stimulated by intriguing fundamental problems involving the interaction of photons, spins, charges and lattices, as well as by the prospects of light-controlled ultrafast magnetic data processing and storage. While research in this field has been primarily focused on the temporal control of magnetic order (*2-4*), optical fields can be simultaneously used for the spatial control of magnetization (*5, 6*). However, the wavelength of visible light restricts our ability to use optical pulses to manipulate and study magnetic phenomena on the nanoscale. Techniques based on near-field optical microscopy have been pursued to circumvent this limitation (*7, 8*) but have so far made limited advances into the nanometer range. Short-wavelength sources in the extreme ultraviolet (EUV) and x-ray ranges offer a prospect of controlling magnetism on a much finer scale than is possible with conventional optical sources. In recent years, EUV and x-rays tuned to absorption edges of elements such as Fe, Ni and Co have been increasingly used to probe and image magnetic phenomena (*9-12*). By contrast, very little work has been done on using short-wavelength radiation to drive magnetic dynamics (*13, 14*).

In this report, we introduce a technique we refer to as transient magnetization gratings, in which coherent femtosecond EUV pulses derived from a free-electron laser (FEL) are used to create transient periodic magnetization patterns with periods of tens of nanometers. The laser-induced transient grating (TG) technique (*15-17*), in which two crossed laser pulses produce a spatially periodic excitation acting as a diffraction grating for the third, time-delayed probe pulse, has been used to study a wide range of phenomena, such as the propagation of acoustic waves (*17, 18*), thermal transport (*19, 20*), carrier and spin dynamics (*21, 22*), charge density waves (*23*) and laser-plasma interaction (*24*). In the studies of magnetism, the TG method was used to excite surface acoustic waves that produced magnetic responses via magnetostriction (*25*). Crossed laser pulses have also been used to record static periodic magnetization patterns (*5*). Recently, this technique has been extended into the EUV range (*26*) and a EUV TG setup is now available at the EIS-TIMER beamline (*27*) at the FERMI FEL (Elettra, Italy) (*28*).

In the previous non-magnetic EUV TG experiments (*26*), the signal was dominated by thermal and acoustic responses, resulting in a periodic mass density modulation in the sample. In this work, we combine the EUV TG excitation with resonant probing at the dichroic M-edge of cobalt, which allows us to create and probe magnetization gratings with a period as small as 44 nm in a CoGd alloy. The magnetization grating appears on a sub-picosecond time scale as the sample is thermally demagnetized at the TG maxima, and then decays on time scales of tens of picoseconds, suggesting that thermal effects provide the leading mechanism for washing away the magnetization grating. We demonstrate that the TG signal in the presence of an external magnetic field greatly exceeds the contributions from the electronic excitation and thermoelastic response. Based on our findings, we propose a number of avenues for using transient magnetization gratings to study ultrafast nanoscale magnetic phenomena.

**Results**

The experiment is schematically depicted in Fig. 1A, B. The sample consists of a thin metal film stack containing 9 nm of $Co_{0.81}Gd_{0.19}$ alloy with perpendicular magnetic anisotropy deposited on a 100 nm-thick $Si_3N_4$ membrane. The film is magnetized to saturation using an external magnetic field normal to the film plane. Two time-coincident 60 fs excitation pulses, crossed at an angle of $2\Theta = 27.6°$, form an interference pattern with a period of $\Lambda = \lambda_{ex}/2\sin\Theta$, where $\lambda_{ex}$ is the excitation wavelength. We used two excitation wavelengths, 41.6 and 20.8 nm, yielding TG periods of 87.2 and 43.6 nm, respectively. The excellent coherence of the FERMI FEL, with EUV pulses which are nearly transform-limited (*27*), ensures a high contrast of the resulting interference pattern within the entire EUV spot size, producing a long-range spatial order of the sample excitation.

Absorption of the excitation radiation by the sample leads to a periodic profile of the electronic temperature: at the maxima of the excitation intensity the temperature rises, resulting in ultrafast demagnetization, whereas at the minima the temperature and the magnetization remain unchanged. Accordingly, the magnetization becomes periodically modulated,

$$M(x) = M_0 \left[1 - \frac{\Delta}{2}(1 - \cos q\, x)\right], \qquad (1)$$

where $M_0$ is the initial magnetization at room temperature, $q = 2\pi/\Lambda$ is the grating wavevector magnitude, and $\Delta$ is the peak-to-peak amplitude of the modulation relative to the initial magnetization.

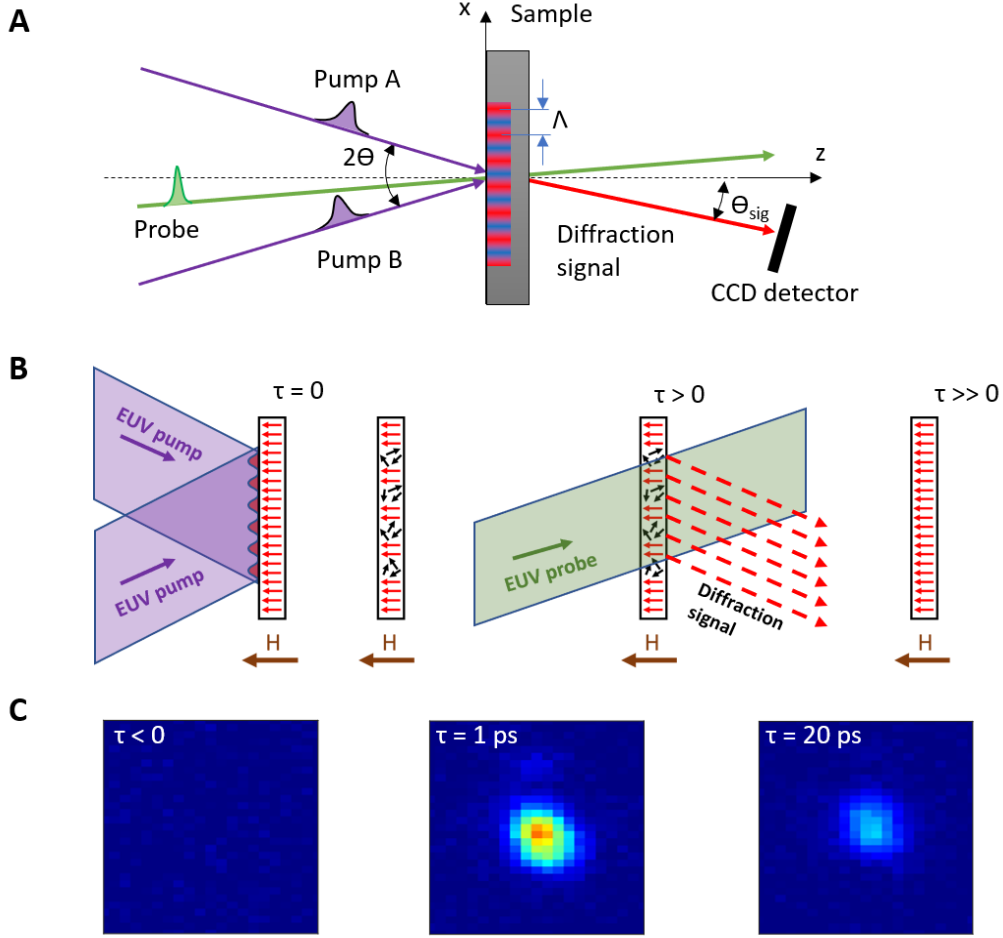

**Fig. 1. Nanoscale transient magnetization grating.** (**A**) Schematic illustration of the transient magnetization grating technique at the nanometer length-scale, based on the interference of the two EUV pump pulses yielding a spatially modulated excitation in the sample with a period Λ. The magnetization grating signal is recorded as a function of the time delay between the pump and probe pulses. (**B**) The sample is kept in a homogenous magnetization state by the external field H, which is larger than the coercive field. Then the two EUV pump pulses generate a nanoscale spatial modulation of the temperature, producing a transient magnetization grating in the sample. The dynamics of such a spatially-periodic excitation are monitored via transient diffraction (dashed arrows) of a time-delayed probe pulse, tuned at a magnetic absorption edge. After each FEL shot the sample is brought into the initial magnetic state by the field H. (**C**) Images of the signal beam, i.e. the transient diffracted beam, collected with a CCD camera and corresponding to different pump-probe delays that are indicated in the individual panels.

We are only interested in the dynamics of the spatially periodic component of the magnetization, given by $M_0 \frac{A}{2} \cos q\, x$.

The dynamics of this magnetization grating are monitored via diffraction of a time-delayed probe pulse, whose wavelength $\lambda_p = 20.8$ nm is tuned to the $M_{2,3}$ edge of Co. The periodic modulation of the magnetization leads to a modulation of the magnetic circular dichroism via the complex refractive indices,

$$n_\pm = \pm(\Delta\delta + i\Delta\beta)\frac{A}{2}\cos q\, x, \qquad (2)$$

where the subscript + (−) refers to the right (left) circular polarization, Δδ and Δβ are the magneto-optical (MO) constants of the sample, which can be estimated from the MO constants of Co (*29*). The probe beam is polarized vertically (out-of-plane in Fig. 1B) and is incident at an angle of $\Theta_p=4.6°$, i.e., close to the sample normal. The modulation of dichroism acts as a depolarizing diffraction grating that produces a horizontally polarized diffracted beam (see Supplementary Material). The diffraction efficiency, i.e., the ratio of the diffracted intensity to the intensity of the transmitted zeroth-order beam, is given by

$$\frac{I_d}{I_0} = \frac{k^2 d^2}{16} \Delta^2 (\Delta\delta^2 + \Delta\beta^2). \qquad (3)$$

where $I_d$ and $I_0$ are the intensities of the 1st order diffracted and transmitted (i.e., zeroth order) beams, respectively, $k_p=2\pi/\lambda_p$ is the probe wavevector, and $d$ is the thickness of the grating. Thus, by measuring the dependence of the diffraction intensity on the pump-probe time delay, we can monitor the dynamics of the amplitude of the magnetization grating $\Delta$. Figure 1C shows representative images of the diffraction spot on a CCD camera, obtained by averaging 2000 FEL shots. (The 50 Hz repetition rate was low enough to ensure that after each shot, the sample cooled down to the background temperature and recovered the initial magnetization state.) At negative time delays, when the probe pulse arrives before the excitation, the sample is uniformly magnetized and no diffraction occurs; at positive delays we see a well-defined diffraction spot indicating the presence of the transient grating; at longer delays the grating decays and the signal fades away. The detection process is selectively sensitive to the spatial Fourier component of the magnetization at the wavevector $q$; it is insensitive to the variation of the average magnetization of the sample and to the modulation at higher spatial harmonics, which may appear due to the nonlinear dependence of magnetization on the excitation fluence

The diffraction signal in Fig. 1C by itself does not prove the existence of the magnetization grating: a TG signal may also result from a modulation of the refractive index caused by electronic excitation and, on a longer time scale, by the density modulation produced by thermal expansion. The diffraction signal from a non-magnetic transient grating would have the same vertical polarization as the incident probe beam, and could be separated from the magnetic TG signal by polarization analysis. Alternatively, one can compare measurements made with and without the external magnetic field, which is the strategy we used here to elucidate the nature of the signals we observed.

Figure 2A shows the dependence of the TG signals on the pump-probe time delay at $\Lambda=87.2$ nm. One can see that in a saturating magnetic field, the signal is much greater than at zero field under similar experimental conditions. Since the non-magnetic contribution to the TG signal is independent of the magnetic field, the signal measured in the saturating field must almost entirely come from the magnetization TG. The magnetic TG signal apparently disappears at H=0. We believe that at the zero field condition the sample is demagnetized to a disordered multidomain state: when the external field is turned off, a large inductance of the coils leads to an oscillating decaying magnetic field, which drives the magnetization along minor hysteresis loops with progressively decreasing remanent magnetization. In this case, the sample consists of alternating up-and-down domains and, while the diffraction intensity does not depend on the sign on the magnetization according to Eq. (3), the electric field in the diffracted beam changes sign when the magnetization sign flips (see Supplementary Material). Consequently, the magnetic TG signal in the direction of the diffracted beam integrates to zero. One can expect that in the saturating field, the signal should not depend on the magnetic field direction. The reason that the signals collected at H=+/-40 mT are not identical lies most likely in the slow drift of the FEL during the roughly three hours' time gap between the two measurements: while the FEL energy stayed about the same, a fluctuation of the beam pointing may have caused a change in the beam overlap at the sample which degraded the signal.

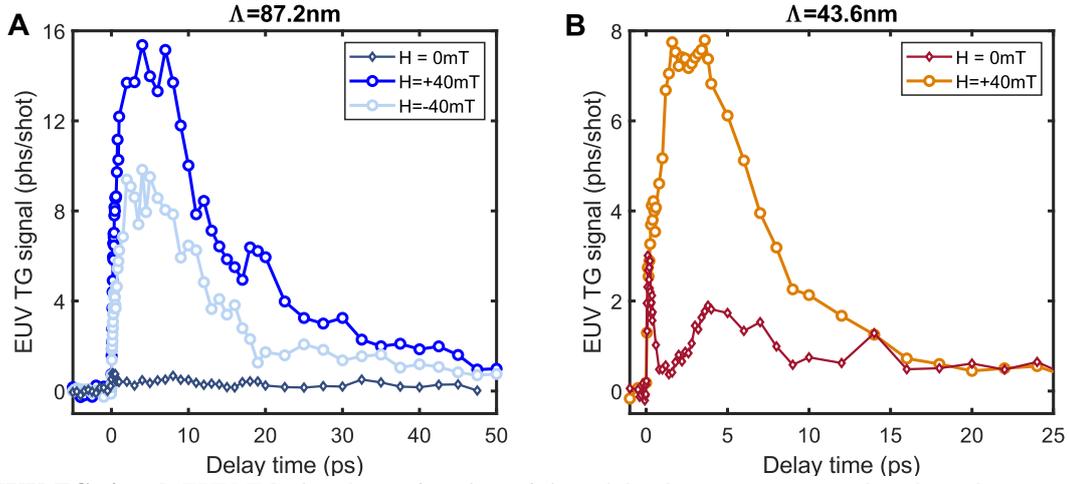

**Fig. 2. EUV TG signal.** EUV TG signal as a function of time delay between pump and probe pulses, recorded without and with positive and negative external magnetic field for TG periodicities $\Lambda$ = 87.2 nm (**A**) and $\Lambda$ = 43.6 nm (**B**); the magnitude of H at 40 mT is larger than the coercive field."

Figure 2B shows the data recorded at $\Lambda$=43.6 nm: in this case the excitation wavelength is tuned to the $M_{2,3}$ edge of Co. The H=0 signal is now more prominent, and one can see that it consists of an initial short peak near t=0, followed by a slower non-monotonic response. The initial peak, shown in more detail in Fig. 3A, must originate from the electronic excitation and subsequent electron-lattice relaxation: its ~50 fs rise time is close to the FEL pulse duration, while the ~0.5 ps decay time is within the expected range for the electron-phonon relaxation in a metal. The slower part of the signal is attributed to the thermoelastic response, which involves acoustic oscillations and thermal decay. Similar thermoelastic signals have been previously observed in EUV TG experiments with a non-resonant probe (*26, 30*); however, in those experiments the electronic peak was barely visible. In our case, the electronic peak is stronger than the thermoelastic response.

We now analyze the behavior of the magnetic responses measured in the saturating field. In both parts of Fig. 2, we see a similar pattern of a fast rise followed by a slower decay. As discussed above, the rise of the signal is caused by the fast demagnetization at the maxima of the excitation grating; we generally expect to see a rise time similar to that reported in pump-probe experiments with optical excitation. Figure 3B provides a detailed view of the initial dynamics. At resonant probe/pumping ($\Lambda$=43.6 nm) we see an initial step-like feature with a rise time of ~50 fs. As becomes apparent from Fig. 3A, this feature results from the "contamination" of the magnetic TG signal by the non-magnetic electronic response. Thus, the signal measured with non-resonant pumping ($\Lambda$=87.2 nm) more faithfully reflects the initial dynamics of the magnetic response. One can see that in the first 250 fs, the signal rises to ~50% of its maximum level (since the signal is proportional to $\Delta^2$, this corresponds to ~70% of the maximum magnetization grating amplitude); this rise time is in line with what was observed with optical excitation on a similar sample when the Co sublattice was probed (*31*). However, thereafter the rise slows down, with the maximum achieved at ~3 ps. The observed complex magnetization behavior has not been seen in any earlier ultrafast magnetization experiments at optical wavelengths. This points towards a different ultrafast magnetic behavior in nanoscale magnetization patterns.

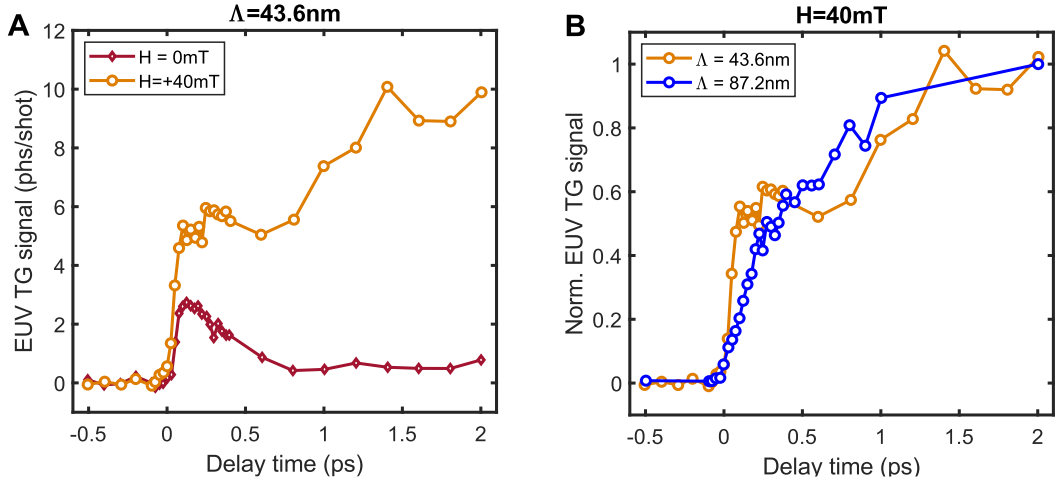

**Fig. 3. EUV TG signal - short time dynamics.** (**A**) EUV TG signal without and with external magnetic field for Λ = 43.6 nm. The peak in the first few hundred femtoseconds corresponds to the electronic response. (**B**) EUV TG signal (normalized to a value of 2 ps) in the first few picoseconds for the two investigated grating periodicities: Λ = 43.6 nm and Λ = 87.2 nm. In both cases the external magnetic field was present (H = 40 mT).

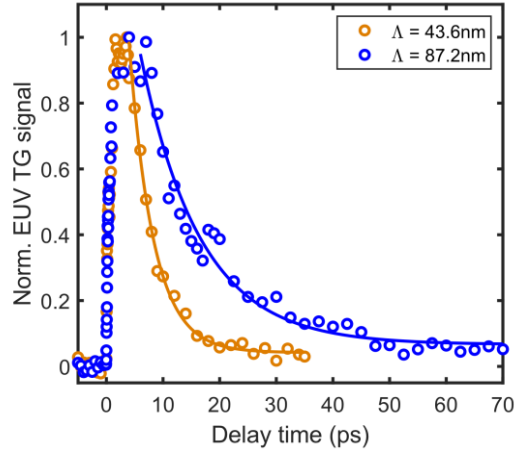

**Fig. 4. Time decay of the EUV TG signal.** Dynamics of the transient magnetization grating signal for Λ = 43.6 nm and Λ = 87.2 nm. In both cases the external magnetic field was H = 40 mT. Solid lines are fits to an empirical model function described in the text.

The decay of the magnetic TG signal occurs on a much slower time scale. As can be seen in Fig. 4, the decay time is longer at a longer TG period: an exponential fit yields a decay time of 10 ps at Λ =43.6 nm and 21 ps at Λ=87.2 nm. The dependence on the TG period indicates that the signal decay involves a transport process. Indeed, on this time scale, the lattice, electronic and spin systems are in thermal equilibrium; as the temperature grating is washed away by thermal diffusion, the magnetization grating also decays. In a one-dimensional thermal diffusion model, the signal decay would be exponential, with a decay time given by $\tau=\Lambda^2/8\pi^2\alpha$, where $\alpha$ is the thermal diffusivity (*15*). In the experiment, the decay time does not follow the $\Lambda^2$ dependence, which is not entirely surprising since we have a multilayer structure and Λ =43.6 nm is not much greater than the total thickness of the metal stack (22 nm). Thus, thermal transport occurs in both in-plane and normal directions, and the one-dimensional model is not expected to be accurate. Still, we can use this model to make an estimate of the effective thermal diffusivity at Λ=87.2 nm, $\alpha_{eff} \sim 5\times10^{-6}$ m$^2$/s. This value is not unreasonable for a stack consisting of ultrathin layers including an alloy (see Supplementary material).

**Table 1. Parameters for both configurations.** Parameters of the pump and probe beams for both configurations.

|  | **Configuration 1** | **Configuration 2** |
| --- | :---: | :---: |
| Grating period $\Lambda$, nm | 43.6 | 87.2 |
| *Pump beam:* | | |
| Wavelength, nm | 20.8 | 41.6 |
| Pulse energy, $\mu$J | 2 x 1.1 | 2 x 2.0 |
| FEL fluence at the sample, mJ/cm$^2$ | 1.5 | 2.5 |
| *Probe beam:* | | |
| Wavelength, nm | 20.8 | 20.8 |
| Pulse energy, $\mu$J | 0.17 | 0.14 |
| FEL fluence at the sample, mJ/cm$^2$ | 0.11 | 0.09 |

**Discussion**

While in the initial EUV TG experiments (*26, 30*), the signal was dominated by the lattice response driven by thermal expansion, the use of a resonant probe makes it possible to employ the EUV TG method to study ultrafast dynamics of photoexcited electronic and spin systems. We have seen that the magnetic TG response probed at the M-edge of Co is in fact much stronger than the thermoelastic response. We estimate that at the signal maximum, the TG efficiency $I_d/I_0$ is ~ 2(5)·10$^{-8}$ for $\Lambda$ = 41.6 nm (87.2 nm). Estimating the MO constants of CoGd from those of Co (*28*) based on the number of Co atoms per unit volume (see Supplementary material), we infer that the magnetization grating amplitude was ~3 – 4% of the initial magnetization value $M_0$. This is not inconsistent with the estimated temperature increase of less than 100 K. It remains to be seen whether stronger EUV-induced demagnetization is possible without destroying the sample in the multi-shot regime.

EUV-driven transient gratings of magnetization open a way for studying ultrafast magnetic phenomena with imposed periodicity on the scale of tens of nanometers, an order of magnitude smaller than achievable with standard domain engineering (*32*), and extending the TG period range down to a few nanometers is within reach with the current setup. Several avenues for further exploration can be identified. The TG technique is well suited for studying spin transport and we believe that the hallmark of spin diffusion will be a dependence of the demagnetization dynamics on the TG period, as the latter is further decreased. A related issue is the smallest magnetization TG period achievable: the smallest region in which the electronic temperature can be defined is determined by the electronic mean free path; likewise, the smallest size of the region in which the magnetization can be defined is limited by the spin mean free path. By measuring electronic, lattice and magnetic TG responses as a function of $\Lambda$, one can determine these lengths for different materials.

The TG technique should enable the excitation of magnons at the TG wavevector; magnon spectroscopy with EUV TG will bridge the gap between Brillouin scattering operating in the wavevector range up to ~0.02 nm$^{-1}$, and neutron / resonant x-ray scattering typically operating above 1 nm$^{-1}$. As a time-domain technique, the TG method offers an additional advantage of being free from instrumental spectral resolution limits. Yet another prospect is pushing the study of magnetoelastic interactions involving surface acoustic waves (*25*) into the hundreds of MHz to THz range. Finally, the TG approach can be used for holographic magnetic recording (*5*) on the nanometer scale. We envision that our work will stimulate further research on nanoscale magnetic transient gratings, with the prospect of further exciting opportunities that cannot be anticipated at this early stage.

## Materials and Methods

### Experimental details

The experiments were performed at the EIS-TIMER end-station. This setup uses wave-front division beam-splitting to produce three FEL beams which are overlapped at the sample. The spot size at the sample is about 0.18 mm$^2$ for the pump beams and 0.16 mm$^2$ for the probe. The excitation pulses have circular polarization (chosen to ensure the most stable FEL operation; since the demagnetization mechanism involves electronic excitation, we do not believe the pump polarization is significant), while the probe pulse is linearly polarized (out of the plane of the drawing in Fig. 1A). The parameters of pump and probe pulses are listed in Table 1. Further details on the optical setup can be found elsewhere (*27*). A holey electromagnet is used to supply a permanent magnetic field orthogonal to the sample surface while still allowing all the EUV beams to reach the sample.

The diffracted probe pulse is detected by a soft x-ray in-vacuum charge-coupled device (CCD) camera (Princeton PI-MTE). The camera has 2048 x 2048 pixels of 13.5 microns size and is triggered by the FEL fast shutter signal allowing for single-pulse recording. The EUV TG signal was determined by integrating the CCD counts in a region of interest around the signal beam and normalized to the cube of the FEL pulse energy. To convert the ADC counts recorded by the CCD, into incoming photons we assumed sensitivity of 5.8 photons per ADC count, which includes the detector quantum efficiency. The FEL fluence for the pump and probe pulses were determined by combining the information from the I0-monitor and EUV spectrometer in the photon diagnostic system upstream of the beamline.

### Sample preparation

The sample structure was the following, counting from the bottom: 100 nm Si$_3$N$_4$ / 4 nm Ta / 5 nm Pt / 9 nm Co$_{0.81}$Gd$_{0.19}$ / 4 nm Ta. The metal stack was deposited on a 100 nm-thick Si$_3$N$_4$ membrane by sputtering at room temperature in 4.9 mTorr of ultrapure Ar, in an ultrahigh vacuum chamber with a base pressure of 6×10$^{-9}$ Torr. The 9 nm-thick Co$_{81}$Gd$_{19}$ film was deposited by co-sputtering from pure Co and Gd targets, with respective deposition rates calibrated by a quartz crystal microbalance to the precision of better than 0.1%. 4 nm-thick Ta/5nm-thick Pt bilayer was used as a buffer layer. The film was protected from oxidation by a 4 nm-thick Ta capping layer. The deposition rates for Ta, Pt, and CoGd were 0.2A/s, 1A/s, and 0.4 A/s, respectively. The magnetization, magnetic anisotropy, and coercive field of the sample were determined by vibrating sample magnetometry at room temperature, and additionally checked by magneto-optic Kerr magnetometry.

**Acknowledgments**

**General**: We gratefully acknowledge the FERMI team for their support throughout the whole project.

**Funding:** D.K. and C.G. acknowledge funding by the Deutsche Forschungsgemeinschaft (DFG) projects GU 535/4-1 and KS 62/1-1. V.U., M.P. and S.B. acknowledge support from the European Research Council, Starting Grant 715452 MAGNETIC-SPEED-LIMIT. The contribution of S.U. was supported by the U.S. Department of Energy, Office of Science, Basic Energy Sciences, under Award # DE-SC0018976. The contribution of the MIT collaborators was supported by the U.S. Department of Energy Award DE-SC0019126.

**Author contributions:** S.B., C.G. and A.M. proposed the experiment. S.U. fabricated and characterized the samples. F.B., M.P. and S.B. designed and realized the magnetic setup. D.K., A.M., V.U., F.B., F.C., A.C., L.F., M.M., R.M., E.P., L.R., H.R. and C.G. performed the experiment. D.K, A.M., R.M., E.P. and C.G. analyzed the data. D.K, A.M., F.B., C.M., K.N., S.B. and C.G. discussed the data and wrote the paper, with input from all co-authors. All authors participated in the discussion and interpreted results. All authors commented on the manuscript.

**Competing interests:** The authors declare that they have no competing interests.

**Data and materials availability:** All data needed to evaluate the conclusions in the paper are present in the paper and/or the Supplementary Materials. Additional data related to this paper may be requested from the authors.

# Supplementary

## S1. Diffraction of the probe beam by a magnetization grating

For the probe beam directed along the sample normal $z$ (we neglect the small incidence angle) and polarized along $y$, the electric field is given by $E_0 \boldsymbol{e}_y exp(ikz - i\omega t)$, where $\boldsymbol{e}_y$ is the unit Jones vector, and $k=2\pi/\lambda$ is the wavevector magnitude. The linearly polarized light can be decomposed into the two circularly polarized components:

$$\boldsymbol{e}_y = \frac{i}{\sqrt{2}}(\boldsymbol{e}_- + \boldsymbol{e}_+), \text{ where } \boldsymbol{e}_\pm = \frac{1}{\sqrt{2}}\begin{pmatrix}1\\ \pm i\end{pmatrix} . \tag{S1}$$

We consider a weak grating of dichroism given by Eq. (2) from the main text,

$$n_\pm = \pm(\Delta\delta + i\Delta\beta)\frac{\Delta}{2}\cos q\, x, \tag{S2}$$

If the diffraction efficiency is low, the diffracted field amplitude for each of the circularly polarized components is given by

$$\boldsymbol{E}_{d\pm} = E_0 \frac{i}{\sqrt{2}} \boldsymbol{e}_\pm \frac{kd\Delta}{4}(\pm i\Delta\delta \mp \Delta\beta)\, exp[-kd(\beta \pm \Delta\beta)], \tag{S3}$$

where $d$ is the thickness of the grating.

Assuming that $kd\Delta\beta<1$, the amplitude of the diffracted electric field is

$$\boldsymbol{E}_d = (\boldsymbol{E}_{d-} + \boldsymbol{E}_{d+}) = E_0 \boldsymbol{e}_x \frac{kd\Delta}{4}(-i\Delta\delta + \Delta\beta)\, exp[-kd\beta], \tag{S4}$$

with the polarization being orthogonal with respect to that of the incident probe beam. The diffraction efficiency, i.e., the ratio of the diffracted intensity to the intensity of the transmitted zeroth-order beam, is given by

$$\frac{I_d}{I_0} = \frac{k^2 d^2}{16}\Delta^2(\Delta\delta^2 + \Delta\beta^2). \tag{S5}$$

## S2. Absorption profile of the multilayered sample (*33*).

The normalized light intensity $I(z)$ at position $z$ in a material is defined by the Poynting vector and can be written as

$$I(z) = n(z)|E(z)|^2, \tag{S6}$$

with $n$ the real part of the complex refractive index $\tilde{n} = n + ik$, where $i$ is the imaginary unit, $k$ the extinction coefficient and $|E(z)|^2$ the electric field intensity. To a first-order approximation, light dissipation in a material is given by

$$I(z) = -\alpha(z)I(z)dz \tag{S7}$$

where the absorption coefficient $\alpha = 4\pi k/\lambda$, where $\lambda$ is the wavelength of the light. From the definition of the absorption ($dA(z) = -dI(z)$), and by combining equations (S6) and (S7) we obtain a general expression for the absorption profile of light in any multilayer structure:

$$dA(z) = -\alpha(z)\tilde{n}(z)|E(z)|^2 dz, \tag{S8}$$

Computations of the electric field intensity $|E(z)|^2$ are based on the application of the Fresnel equations and use of the atomic scattering factors (*34*) for the refractive indices of the layers (Table S1).

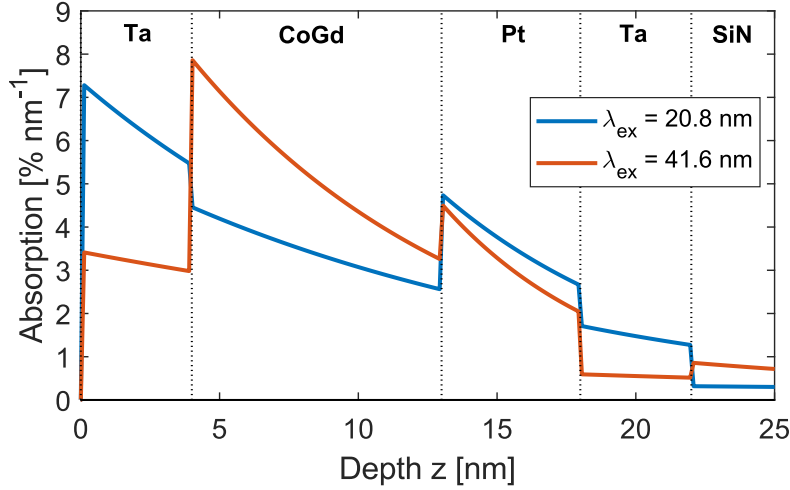

**Fig. S1. Absorption profiles in the sample.** Calculated absorption profiles as a function of depth inside the multilayer for the pump beam with wavelengths $\lambda_{ex}$ = 20.8 and 41.6 nm.

**Table S1.** Parameters of multilayer used for the calculation of the absorption profiles.

| Layer | d, nm | 20.8 nm | | 41.6 nm | |
|---|---|---|---|---|---|
| | | $n$ | $k$ | $n$ | $k$ |
| Ta | 4 | 0.853 | 0.120 | 0.898 | 0.114 |
| CoGd | 9 | 0.973 | 0.099 | 0.832 | 0.315 |
| Pt | 5 | 0.918 | 0.188 | 0.746 | 0.515 |
| Ta | 4 | 0.853 | 0.120 | 0.898 | 0.114 |
| SiN | 100 | 0.925 | 0.030 | 0.747 | 0.194 |

### S3. Magneto-optical (MO) constants for CoGd alloy

The optical properties of $Co_{0.81}Gd_{0.19}$ are

$$\delta_{CoGd}(\beta_{CoGd}) = \left(\rho_m N_a r_0 \frac{\lambda^2}{2\pi}\right) \frac{0.81 f_{Co} + 0.19 f_{Gd}}{A_r(Co_{0.81}Gd_{0.19})}$$
$$= \left(\rho_m N_a r_0 \frac{\lambda^2}{2\pi}\right) \frac{0.615 f_{Co}}{A_r(Co)} + \left(\rho_m N_a r_0 \frac{\lambda^2}{2\pi}\right) \frac{0.385 f_{Gd}}{A_r(Gd)} =$$
$$= 0.59 \left(\rho_m(Co) N_a r_0 \frac{\lambda^2}{2\pi}\right) \frac{f_{Co}}{A_r(Co)} + 0.41 \left(\rho_m(Gd) N_a r_0 \frac{\lambda^2}{2\pi}\right) \frac{f_{Gd}}{A_r(Gd)}$$
$$= 0.59 \delta_{Co}(\beta_{Co}) + 0.41 \delta_{Gd}(\beta_{Gd})$$

Assuming that the MO constants at the Co M edge are determined by the number of Co atoms per unit volume, the MO values for Co (*29*) should be multiplied by 0.59 to get the MO constants of CoGd.